\documentclass[11pt]{article}
\usepackage{epsfig,dsfont}
\usepackage{geometry}
\usepackage[T1]{fontenc}
\usepackage[utf8]{inputenc}
\usepackage[english]{babel}
\usepackage{amsmath}
\usepackage{amsfonts}  
\usepackage{epsfig}
\usepackage{calc}
\usepackage{graphicx}
\usepackage{xcolor} 
\usepackage{cite}

\begin{document}  
\sffamily

\thispagestyle{empty}
\vspace*{15mm}

\begin{center}

{\LARGE
Remarks on the construction of worm algorithms  for \vskip3mm 
lattice field theories in worldline representation}
\vskip25mm
Mario Giuliani, Christof Gattringer
\vskip5mm
University of Graz \\
Institute for Physics\\ 
A-8010 Graz, Austria 
\end{center}
\vskip15mm

\begin{abstract}
We introduce a generalized worldline model where the partition function is a sum over configurations of a conserved flux 
on a $d$-dimensional lattice. The weights for the configurations of the corresponding worldlines have factors living on the 
links of the lattice, as well as terms which live on the sites $x$ and depend on all fluxes attached to $x$. 
The model represents a general class of worldline systems, among them the dual representation of 
the relativistic Bose gas at finite density. We construct a suitable worm algorithm and show how to correctly distribute 
the site weights in the various Metropolis probabilities that determine the worm. We analyze the algorithm in detail 
and give a proof of detailed balance. Our algorithm admits
the introduction of an amplitude parameter $A$ that can be chosen freely. Using a numerical simulation of the relativistic Bose gas 
we demonstrate that $A$ allows one to influence the starting and terminating
probabilities and thus the average length and the efficiency of the worm. 
\end{abstract}

\setcounter{page}0
\newpage

\section{Introduction}
Since its introduction in \cite{worm}, the worm algorithm and its generalizations have played an increasing role for Monte Carlo 
simulation of systems with constraints. An interesting class of systems where worm algorithms have been applied are lattice 
field theories (see \cite{endres} --\cite{katz} for 
examples related to the systems studied here). Part of that interest comes from the fact that 
the complex action problem, caused by a non-vanishing chemical potential $\mu$, can in some cases be solved by exactly
rewriting the system in terms of new, so-called dual variables. The complex action problem is the fact that in 
the conventional representation in terms of field variables
the action $S$ of many field theories becomes complex at finite $\mu$ and the Boltzmann factor $\exp(-S)$ cannot be used 
as a probability in a Monte Carlo simulation. If a suitable dual representation can be found, the new form of the partition sum
has only real and positive contributions, such that a simulation directly in terms of the dual variables solves the complex action
problem. The dual variables for matter fields are closed loops of conserved flux, often referred to as worldlines, while gauge fields
are represented as 2-dimensional world sheets (not considered here).

The fact that the dual variables for matter fields are loops of conserved flux makes the worm algorithm the perfect tool for
updating these worldlines. However, the structure of the dual representation of lattice field theories is often more complicated 
than the one encountered for spin systems used for the original formulation of the worm algorithm \cite{worm}:  
Often new weight factors appear that live on the sites $x$ of the lattice (not only on the links as in the case of most spin systems). 
Furthermore, these additional weight factors depend on all the flux variables connected with $x$, and the construction of 
a suitable worm algorithm has to include these weight factors correctly in the probabilities for the
Metropolis acceptance steps defining the worm. 
Although worm-type algorithms have been used for updating such more general systems 
\cite{endres} --\cite{katz}, surprisingly little theoretical background, such as proofs of the detailed balance condition 
(or any other condition ensuring the correct distribution of the configurations) can be found in the literature\footnote{An interesting 
exception is \cite{Rindlisbacher} where detailed balance is proven for a generalized worm algorithm for the complex scalar field 
with external charges.}. The goal of this paper is to address this issue for a certain
general class of systems and to put the use of worm algorithms on a more sound theoretical basis for these models. 
 
A simple example for the type of systems we consider is given by the worldline representation of the Relativistic Bose Gas (RBG). 
In the conventional form the RBG on the lattice is described by the action
\begin{equation}
S_{RBG} \; = \; \sum_x \!\left( \eta |\phi_x|^2 + \lambda |\phi_x|^4 - 
\sum_{\nu = 1}^4 \left[ e^{\mu \, \delta_{\nu,4} } \phi_x^\star \phi_{x+\widehat{\nu}}  \, + \,
e^{-\mu \, \delta_{\nu,4} } \phi_{x+\widehat{\nu}}^\star \, \phi_{x}  \right]  \right) \; ,
\label{action}
\end{equation}
where the first sum runs over the sites $x$ of a 4-dimensional lattice, the second sum is over the four 
Euclidean directions $\nu = 1,2,3,4$, and $\widehat{\nu}$ denotes the unit vector in $\nu$-direction. 
In the conventional representation the degrees of freedom are the complex valued field variables $\phi_x$ at the
sites $x$ of the lattice.  $\eta$ denotes the combination $8 + m^2$, where $m$ is the bare mass parameter. 
The coupling of the quartic self-interaction is denoted by $\lambda$ and the chemical potential by $\mu$. 
The partition sum $Z$ is obtained by integrating the Boltzmann factor $e^{-S_{RBG}}$ over all field configurations, i.e., 
$Z_{RBG} = \int\!  D[\phi] \, e^{-S_{RBG}}$, with the product measure $D[\phi] = \prod_x \int_\mathds{C} d \phi_x / 2\pi$. It is obvious
that for $\mu \neq 0$ the action (\ref{action}) has a non-zero imaginary part and the RBG thus 
has a complex action problem in the conventional representation. 
The RBG is an important model system which has been studied not only
with dual techniques, but also with other numerical approaches to the complex action problem (see, e.g., 
\cite{complexlangevin} --\cite{thimble}).

The RBG has a dual representation in terms of worldlines which completely 
solves the complex action problem (for the derivation of the dual form used here see \cite{bosedual}). 
In the dual representation the partition sum is given by
\begin{eqnarray}
\hspace*{-7mm} 
&& \hspace*{12mm} Z_{RBG} \; = \; \sum_{\{l\}} \sum_{\{k\}}  
\left[ \prod_x \delta \! \left( \, \sum_{\sigma=1}^d \big[ k_{x,\sigma}  -  k_{x-\widehat{\sigma},\sigma}  \big] \! \right) \right] \; \times
\label{bosedual} \\
\hspace*{-7mm}  &&
\left[ \, \prod_{x,\nu}\! \frac{e^{\,\mu \, \delta_{\nu,4} \, k_{x,4} }}{(|k_{x,\nu}| + l_{x,\nu})! \, l_{x,\nu}!} \right] \!\!
\left[  \prod_x  
F\!\left( \, \sum_{\sigma=1}^d \big[ |k_{x,\sigma}| +  |k_{x-\widehat{\sigma},\sigma}| + 
2( l_{x,\sigma} + l_{x-\widehat{\sigma},\sigma}) \big]  \right) \right]\! . 
\nonumber
\end{eqnarray}
Here the dynamical degrees of freedom are represented by two sets of variables living on the links of the lattice: 
The flux variables $k_{x,\nu} \in \mathds{Z}$ and the auxiliary variables $l_{x,\nu} \in \mathds{N}_0$. The partition sum 
is a sum over all configurations of both variables. While the $l_{x,\nu}$ are unconstrained, the flux  
$k_{x,\nu}$ is conserved, i.e., it obeys the discretized version of the vanishing divergence condition,
\begin{equation}
\label{fluxconservation}
\sum_{\sigma = 1}^d  \nabla_\sigma \, k_{x,\sigma} \;  \equiv \; 
\sum_{\sigma = 1}^d ( k_{x,\sigma} - k_{x-\hat{\sigma},\sigma} ) \; = \; 0 \; \; \; \forall \, x \; ,
\end{equation}
which is implemented by the product of Kronecker deltas
in the first line of (\ref{bosedual})  (we use the notation $\delta(n) \equiv \delta_{n,0}$). The flux conservation implies that the admissible
configurations of $k_{x,\nu}$ are closed worldlines. 

The partition sum has two types of weight factors, both displayed between square brackets in 
the second line of (\ref{bosedual}): The first one lives on the links $(x,\nu)$ and depends only on the dual variables on that 
link. The second type of weight factors lives on the sites $x$ of the lattice and depends on all dual variables that live on links
that are attached to $x$. The function $F(n)$ is given by the integral $F(n) = \int_0^\infty dr \, r^{n+1} \,  e^{-\eta r^2 - \lambda r^4}$.
Obviously all weight factors in (\ref{bosedual}) are real and positive also for $\mu \neq 0$ and the dual formulation 
in terms of worldlines of flux solves the complex action problem. 

While the unconstrained variables $l_{x,\nu}$ can easily be updated with standard Monte Carlo techniques, for the 
flux variables $k_{x,\nu}$ the constraints (\ref{fluxconservation}) have to be taken into account, and the strategy of the
worm algorithm is the method of choice in such a situation. However, as already outlined in the initial paragraphs of this introduction, 
the challenge
here is that in addition to the weight factors living on the links (which are the only ones appearing in the original formulation 
of the worm algorithm \cite{worm}), in a model such as (\ref{bosedual}) we also have weights
that live on the sites $x$ and depend on all flux variables connected with $x$. In such a case there are many possible ways
to include these additional weight factors in the Metropolis acceptance probability for a new step of the worm. 
Compared to the standard worm applications with weights only on the links, in the general case considered here, it 
is a non-trivial challenge to construct a correct Metropolis step for the worm and to prove detailed balance. 

In this paper we study a worldline model which generalizes the dual representation (\ref{bosedual}) 
of the RBG. That model has local link weights, as well as local site weights, and is a prototype for a large class
of worldline models that can be updated with a worm algorithm. We give an explicit construction of a worm algorithm and provide 
a detailed analysis, and in particular prove that the algorithm obeys the detailed balance condition. This proof closes a gap 
in the understanding of worm algorithms for more general applications and puts worldline simulations of such systems on a sound
theoretical basis.

Furthermore we show that a free parameter that can be introduced in our algorithm which  
allows one to influence properties of the worms. 
Using a numerical study of the relativistic Bose gas, we demonstrate that this parameter changes the probability for starting 
and terminating the worm and thus can be used to optimize the length and the numbers of fluxes that are changed per worm.
 
\section{The general worldline model}
The model we consider has a single integer valued conserved flux $k_{x,\nu} \in \mathds{Z}$ living on the links of the
$d$-dimensional lattice 
$\Lambda \equiv \{ x = (x_1, x_2 \, ... \, x_d) \, | \, x_\nu = 1,2 \, ... \, N_\nu , \, \nu = 1,2 \, ... \, d \}$, which we assume 
to have periodic boundary conditions in all directions $\nu = 1,2 \, ... \,d$. The links of the lattice are denoted as $(x,\nu)$ with 
$x \in \Lambda$ and $\nu = 1,2 \, ... \, d$, i.e., they are the oriented connection between the lattice sites $x$ and $x + \hat{\nu}$. 
For later use we introduce also the alternative labeling of links with 
negative direction indices $\nu$ and the corresponding notation for the flux variables,
\begin{equation} 
(x,-\nu) \; \equiv \; (x-\hat{\nu},\nu)\; , \; \, \nu = 1,2 \, ... \, d \qquad \mbox{and} \qquad 
k_{x,-\nu} \; \equiv \; k_{x-\hat{\nu},\nu} \; , \; \,  \nu = 1,2 \, ... \, d \; .
\label{negativesteps}
\end{equation}
The flux $k_{x,\nu}$ is conserved, i.e., it obeys the discretized version of the vanishing divergence condition
(\ref{fluxconservation}). The partition function of the model we consider is given by
\begin{equation}
Z  \; = \; \sum_{\{k\}} \! \left[ \prod_x \! \delta \! \left( \! \sum_\sigma ( k_{x,\sigma} - k_{x-\hat{\sigma},\sigma} ) \! \right) \!\right]  
\left[ \, \prod_{x,\nu} L_{x,\nu} (k_{x,\nu}) \right] 
\left[ \prod_x S_x( \{k_{x,\bullet} \}) \right] 
\; .
\label{Zflux}
\end{equation}
The partition function is a sum $\sum_{\{k\}}$ over all configurations of the flux variables $k_{x,\mu}\in \mathds{Z}$. 
As in the dual RBG partition sum (\ref{bosedual}), the flux conservation condition (\ref{fluxconservation}) 
is implemented by the product over the corresponding Kronecker deltas, i.e., the first term in the sum (\ref{Zflux}), and
the admissible configurations of $k_{x,\mu}$ thus are closed worldlines of flux.

The other two factors under the sum (\ref{Zflux}) are two types of weight factors for
the configurations: The first type of weight factors is located on the links $(x,\nu)$ of the lattice and is formulated 
as a product over the real and positive link weight factors $L_{x,\nu}(k_{x,\nu}) \in \mathds{R}_+$. The link weight factors 
$L_{x,\nu}(k_{x,\nu})$ depend only on the flux variable $k_{x,\nu}$ sitting on the link $(x,\nu)$. Furthermore they can be different for 
different links, i.e., they can have an explicit dependence on $(x,\nu)$, which can, e.g., be used to take into account a second 
dual variable, such as the auxiliary variable $l_{x,\nu}$ of the RBG (\ref{bosedual}). Also a chemical potential 
which gives a different weight to fluxes in positive and negative time direction can be implemented via the explicit dependence 
of the link weight factors $L_{x,\nu}(k_{x,\nu})$ on $(x,\nu)$.

The other type of weight factors is located on the sites $x$ of the lattice and is formulated 
as a product over the real and positive site weight factors $S_x( \{k_{x,\bullet} \})  \in \mathds{R}_+$.
These depend on all flux variables that live on links attached to the site $x$, i.e., on the set of links 
$\{k_{x,\bullet} \} \equiv \{k_{x,\sigma}, \sigma = \pm 1 \, ... \, \pm d \}$.
Also here we allow for a local dependence on the site $x$, which can, e.g., again be used to take into account the second 
dual variable $l_{x,\nu}$ of the RBG (\ref{bosedual}). 

For illustration we note that for representing the flux representation of the RBG for a fixed background configuration 
of the $l_{x,\nu}$ variables one chooses the link and site weight factors
\begin{eqnarray}
\hspace*{-5mm} && L_{x,\nu}(k_{x,\nu}) \; = \; 
\frac{e^{\,\mu \, \delta_{\nu,4} \, k_{x,\nu} }}{(|k_{x,\nu}| + l_{x,\nu})! \, l_{x,\nu}!} \; \, ,
\label{LandS}
\\
\hspace*{-5mm}  && S_x( \{k_{x,\bullet} \})  \; = \; 
F\!\left( \, \sum_{\sigma = 1}^d 
\big[ |k_{x,\sigma}| +  |k_{x-\widehat{\sigma},\sigma}| + 2( l_{x,\sigma} + l_{x-\widehat{\sigma},\sigma}) \big]  \! \right)  \; .
\nonumber
\end{eqnarray}
The dual partition function (\ref{bosedual}) of the RBG is then obtained as $Z_{RGB} = \sum_{\{ l \}} Z$ with $Z$ given by 
(\ref{Zflux}), using the weights (\ref{LandS}). As already mentioned, the update of the unconstrained variables $l_{x,\mu}$
of the RBG can be trivially implemented with standard methods and is not considered here.

We conclude the discussion of the generalized model described by (\ref{Zflux}) with a few comments about observables. In a dual 
representation also the observables need to be represented in terms of the flux variables. For bulk observables, which are obtained
as derivatives of the free energy with respect to parameters of the theory, they assume the form of moments of sums of the flux 
variables. Often also interesting topological properties of the observables become evident in the dual representation: In particular
conserved charges are represented as a winding number of the corresponding worldlines about the 
compactified time direction. Finally it 
is possible to study also $n$-point functions in the dual formalism: They are represented by open strings of flux connecting the space 
time points where the fields of the $n$-point function are placed (see, e.g., \cite{bosedual,kloiber,Rindlisbacher} 
for details of defining the observables for the RBG (\ref{bosedual})).
 
\section{Definition of the worm algorithm}
Having introduced the general worldline model, we now come to the definition of the update with a worm
algorithm. The key idea of the worm algorithm is to violate the constraint (\ref{fluxconservation}) at the endpoints of a
randomly chosen starting link $(x_0,\nu_0)$ by changing the corresponding flux to $k_{x_0,\nu_0} + \, \mbox{sign}(\nu_0) \Delta$,
where $\Delta$ is the flux increment of the worm, with $\Delta \in \{-1,1\}$ chosen randomly. 
Subsequently the worm propagates the defect at $x_1 \equiv  x_0 + \hat{\nu}_0$ 
across the lattice by making random choices for the directions 
$\nu_j \in \{\pm1, \, ... \, \pm d\}$ and so determines the sites $x_{j} \equiv x_{j-1} + \hat{\nu}_{j-1}$ and the 
links $(x_j,\nu_j)$ along which it moves in steps labelled by $j = 0,1,2 \, ...\, $. 
In each step it proposes to change the flux variable of that link according to 
\begin{equation}
k_{x_j,\nu_j} \; \rightarrow \; k^{trial}_{x_j,\nu_j} \; = \; k_{x_j,\nu_j} + \,\mbox{sign}(\nu_j)\, \Delta \; .
\label{fluxchange}
\end{equation} 
Each step of the worm is accepted with a local
Metropolis decision. The worm terminates when a final step is accepted, where it reaches its starting point $x_0$. 
Then the defects are healed and we have a new configuration where all flux variables on links along a closed loop 
${\cal L}$ were changed: 
If $C_k$ denotes the starting configuration given by flux variables $k_{x,\nu}$ obeying (\ref{fluxconservation}), then the worm changes 
$C_k$ into a new configuration $\widetilde{C}_k$, where 
the fluxes on the links $(x,\nu)$ of the closed loop ${\cal L}$ are given by 
\begin{equation} 
\widetilde{k}_{x,\nu} \; = \; k_{x,\nu} + \,\mbox{sign}(\nu)\, \Delta \qquad \forall \, (x,\nu) \in {\cal L} \; .
\label{ktilde}
\end{equation}
All other flux variables remain unchanged. The new configuration $\widetilde{C}_k$ also obeys the constraints 
(\ref{fluxconservation}), since the rule (\ref{ktilde}) ensures that the total change of flux is also conserved at each site $x$. It 
is straightforward to see that any admissible configuration of fluxes that obeys (\ref{fluxconservation}) can be transferred by 
finitely many worms into any other admissible configuration, such that the worm algorithm is ergodic.

It is convenient to introduce the following notation for a worm $w$ of length $n$,
\begin{equation}
w \; = \; (   x_0, \nu_0,  \Delta, \nu_1, \, ... \, \nu_{n-1}) \; ,
\label{wormnotation}
\end{equation}
i.e., we characterize the worm by its starting link, given by the starting site 
 $x_0 \in \Lambda$ and the starting direction $\nu_0  \in \{\pm1, \, ... \, \pm d\}$, by the
 flux increment $\Delta \in \{-1,1\}$, and by the choices it makes for 
the directions $\nu_j \in \{\pm1, \, ... \, \pm d\}$ in each step $j = 0,1,2 \, ...\,$. 
The total number of steps is given by $n$, and the worm visits the
sites $x_j = x_0 + \sum_{i= 0}^{j-1} \widehat{\nu}_i$ in the steps
labelled with $j = 0,1 \, ... \, n-1$. Note that the condition that the worm closes gives rise to
$x_n \equiv x_0 + \sum_{i= 0}^{n-1} \widehat{\nu}_i \;\, {!\, \atop}\!\!\!\!\!\!= x_0$ and thus to $\sum_{i= 0}^{n-1} \widehat{\nu}_i \equiv
\sum_{\nu = 1}^d \widehat{\nu} \, N_\nu \, q_\nu$, where $N_\nu$ is the extent of the lattice in direction $\nu$  and $q_\nu$ 
is the winding number of the worm around that direction (note that due to the periodic boundary conditions the lattice is a 
(hyper-) torus).

\begin{figure}[t!]
\begin{center}
\includegraphics[width=15cm,clip]{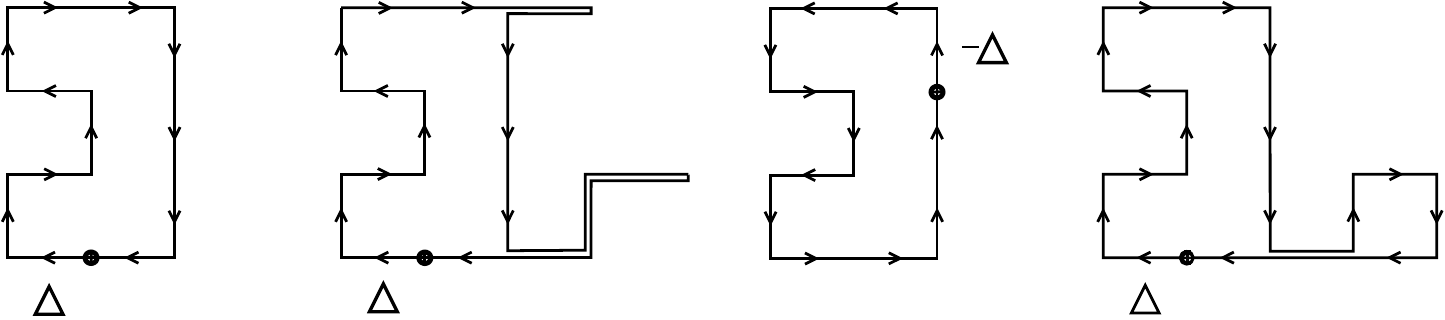}
\end{center}
\vspace{-4mm}
\caption{Examples of worms that change the flux along the same closed loop but differ by dangling ends (first and second worm),
the starting site of the loop, and the orientation and flux increment $\Delta$ (third worm). The starting points of the worms are 
marked with a circle, the orientation with arrows and we denote the flux increment next to the first step of the worm. The 
example on the rhs.~shows that the loop ${\cal L}$ where a worm changes the flux can consist of several disconnected pieces.}
\label{wormcontours}
\end{figure}

We remark that the worm can also retrace one or more steps, i.e., $\nu_{j+1} = -\nu_{j}$ is a legitimate choice and in this 
case the flux variable $k_{x_j,\nu_j}$ remains unchanged due to the rule (\ref{fluxchange}) and the convention (\ref{negativesteps}). 
This implies that a worm can have dangling dead ends, where the 
flux remains unchanged, which are not part of the closed loop ${\cal L}$
where the flux is actually altered when the worm is completed. In the two diagrams on the lhs.\ of Fig.~\ref{wormcontours} 
we show two examples of worms that change the flux along the same closed loop, but differ by several dangling dead ends. 
Note that, although they change the flux along the same loop ${\cal L}$, worms with dangling dead ends have a different 
length. In addition to worms with dangling dead ends also other worms give rise to the same loop with changed flux: 
The worm can start at a different site of the loop or run through the loop in opposite direction with flux increment 
$-\Delta$ instead of $\Delta$. The third diagram in Fig.~\ref{wormcontours} is an example of such a worm. We conclude that there are 
infinitely many worms that lead from a given configuration to another configuration where the flux is changed along a closed 
loop ${\cal L}$. For proving detailed balance also these loops with dangling dead ends will have to be 
considered. We finally point out that the loop 
${\cal L}$ where the flux is changed is not necessarily a single loop, but can consist 
of disconnected pieces. An example of such a loop is shown in the rhs.\ plot of Fig.~\ref{wormcontours}.

We now describe the general form of the worm algorithm using pseudocode. The code has two parts: The part for
starting the worm and the pseudocode for running and terminating the worm. Metropolis decisions determine the acceptance
of the various steps of the worm with the Metropolis probability {\tt min}$\{ \, \rho \, , \, 1 \}$, where $\rho$ is referred to 
as the Metropolis ratio. We use three different Metropolis ratios: $\rho^{\,S}$ is used for starting the worm, $\rho^{\,R}$ for
running the worm, and $\rho^{\,T}$ for terminating the worm. The corresponding mathematical expressions will be given later. In our
code {\tt rand()} denotes a random number generator that provides random numbers uniformly distributed 
in the interval $[0,1)$.

\vskip4mm

\noindent
\underline{{\bf Pseudocode for starting the worm:}}
\vskip2mm
{\tt 

\vskip2mm

randomly select a starting site $x_0$ and a starting direction $\nu_0 \in \{ \pm 1, \pm 2 \, ... \, \pm d \}$

randomly select a flux increment $\Delta \in \{-1,1\}$

\vskip1mm

$x_1^{trial} = x_0 + \widehat{\nu}_0$

\vskip1mm

$k_{x_0,\nu_0}^{trial}  = k_{x_0,\nu_0} + \, \mbox{sign}(\nu_0) \, \Delta$

\vskip2mm

compute $\rho_{x_0,\nu_0}^{\, S} $

\vskip1mm

if rand() < min $\big\{ \, \rho_{x_0,\nu_0}^{\, S} \, , \, 1 \, \big\}$ then

\vskip1mm

\hspace*{5mm} $x_1 \; \leftarrow \; x_1^{trial}$

\hspace*{5mm} $k_{x_0,\nu_0} \; \leftarrow \; k_{x_0,\nu_0}^{trial}$

\hspace*{5mm} run the worm until termination

\vskip1mm

end if
}

\newpage
\noindent
\underline{{\bf Pseudocode for running and terminating the worm:}}
\vskip5mm
{\tt 

while worm is not terminated

\vskip2mm

\hspace*{5mm} randomly select a direction $\nu_j \in \{ \pm 1, \pm 2 \, ... \, \pm d \}$

\vskip1mm

\hspace*{5mm} $x_{j+1}^{trial} = x_j + \widehat{\nu}_j$

\vskip1mm

\hspace*{5mm} $k_{x_j,\nu_j}^{trial}  = k_{x_j,\nu_j} + \, \mbox{sign}(\nu_j) \, \Delta$

\vskip2mm

\hspace*{5mm} if $x_{j+1}^{trial} \neq x_0$ then

\vskip1mm

\hspace*{12mm} compute $\rho_{x_j,\nu_j}^{\, R}$

\hspace*{12mm} if rand() < min $\big\{ \, \rho_{x_j,\nu_j}^{\, R} \, , \, 1 \, \big\}$ 

\vskip1mm

\hspace*{19mm} $x_{j+1} \; \leftarrow \; x_{j+1}^{trial}$

\hspace*{19mm} $k_{x_j,\nu_j} \; \leftarrow \; k_{x_j,\nu_j}^{trial}$

\vskip1mm

\hspace*{12mm} end if

\vskip1mm

\hspace*{5mm} else

\hspace*{12mm} compute $\rho_{x_j,\nu_j}^{\,T} $

\vskip2mm

\hspace*{12mm} if rand() < min $\big\{ \, \rho_{x_j,\nu_j}^{\,T}  \, , \, 1 \, \big\}$ then

\vskip1mm

\hspace*{19mm} $n \leftarrow j+1$

\hspace*{19mm} $k_{x_{n-1},\nu_{n-1}} \; \leftarrow \; k_{x_{n-1},\nu_{n-1}}^{trial}$

\hspace*{19mm} worm is terminated

\vskip1mm

\hspace*{12mm} end if

\hspace*{5mm} end if

\vskip2mm

end while

}

\vskip10mm

We stress that if the worm does not start, then the unchanged configuration $C_k$ is also the next configuration in the Markov
chain of configurations. 

As is obvious from the pseudocode, once it has started the worm propagates the defect introduced at the 
endpoint $x_1 = x_0 + \hat{\nu}_0$ of the starting link, by proposing to add new links $(x_j,\nu_j)$ and 
changing the corresponding flux to $k_{x_j,\nu_j}^{trial}  = k_{x_j,\nu_j} + \, \mbox{sign}(\nu_j) \, \Delta$. This trial flux is accepted with
the Metropolis probability {\tt min}$\{ \, \rho \, , \, 1 \}$ until the worm reaches its starting site $x_0$ and accepts the terminating step. 
The corresponding Metropolis ratios $\rho^{\, S}$, $\rho^{\, R}$ and $\rho^{\, T}$ now have to be chosen such that the configurations 
$C_k$ of the flux variables $k_{x,\nu}$ have the correct distribution according to their weights in the partition sum. 
From the partition sum (\ref{Zflux}) we read off the probability weight $W(C_k)$ for an admissible configuration $C_k$, 
\begin{equation}
W(C_k)  \; = \; \frac{1}{Z} \,
\left[ \, \prod_{x,\nu} L_{x,\nu} (k_{x,\nu}) \right] 
\left[ \prod_x S_x( \{k_{x,\bullet} \}) \right] 
\; .
\label{PC}
\end{equation}
The worm algorithm changes a configuration $C_k$ of the flux variables into a new configuration $\widetilde{C}_k$ 
which differs from $C_k$ by a closed loop ${\cal L}$ along which the flux has been changed. As we have already discussed,
there are infinitely many different worms that change $C_k$ into $\widetilde{C}_k$. These are worms which differ by 
dangling dead ends, by their starting site on the loop or by their orientation 
(using a flux increment $-\Delta$ instead of $\Delta$). All these worms contribute
to the total transition probability $P(C_k \rightarrow \widetilde{C}_k)$ from ${C}_k$ to $\widetilde{C}_k$ and will have to be 
taken into account in our analysis. 

A key property for
generating configurations $C_k$ with distribution (\ref{PC}) is the detailed balance condition 
(see, e.g., \cite{landaubinder} for a standard textbook, or \cite{luscher} for a review in the context of lattice field theory),
\begin{equation}
\frac{P(C_k \rightarrow \widetilde{C}_k)}{P(\widetilde{C}_k \rightarrow C_k)} \; = \; \frac{W(\widetilde{C}_k)}{W(C_k)} \; .
\label{detailedbalance}
\end{equation}
Note that in addition to (\ref{detailedbalance}), the transition probability 
$P(C_k \rightarrow \widetilde{C}_k)$ also needs to be defined as a proper, normalized probability,
i.e.,  $P (C_k \rightarrow \widetilde{C}_k) \geq 0 ~ \forall ~ \widetilde{C}_k $ and  
$\sum_{\widetilde{C}_k} P (C_k \rightarrow \widetilde{C}_k) = 1$.
Together with detailed balance this implies the fixed point condition 
$\sum_{{C}_k} W(C_k) P (C_k \rightarrow \widetilde{C}_k)  = W(\widetilde{C}_k)$ 
of the Monte Carlo process. This is a sufficient condition for the process to generate the correct distribution 
$W(C_k)$.

The key step for constructing a worm algorithm that obeys detailed balance 
for the general worldline model is to correctly distribute the 
link and site weight factors $L_{x,\nu} (k_{x,\nu})$ and $S_x( \{k_{x,\bullet} \})$ 
in the definitions for the Metropolis ratios $\rho^{\, S}$, $\rho^{\, R}$ and $\rho^{\, T}$  
that determine the worm. A correct assignment, which will allow us to prove detailed balance 
in the next section and thus determines the correct 
worm algorithm for the general worldline model, is given by
\begin{eqnarray} 
\rho_{x_0,\nu_0}^{\,S} & = & \frac{A}{
S_{x_0}( \{k_{x_0,\bullet} \}) \; S_{x_1}( \{k_{x_1,\bullet} \})} \;
\frac{ L_{x_0,\nu_0}(k_{x_0,\nu_0}^{trial})}{L_{x_0,\nu_0}(k_{x_0,\nu_0})} \; ,
\label{rhoS}
\\
& & 
\nonumber \\
\rho_{x_j,\nu_j}^{\,R} & = &  \frac{
S_{x_j} ( \{ k^{new}_{x_j,\bullet} \}) }{ 
S_{x_{j+1}}( \{k_{x_{j+1},\bullet} \}) } \;
\frac{ L_{x_j,\nu_j}(k_{x_j,\nu_j}^{trial})}{L_{x_j,\nu_j}(k_{x_j,\nu_j})} \; ,
\label{rhoR}
\\
& & 
\nonumber \\
\rho_{x_{j},\nu_{j}}^{\,T} \Big|_{j \,=\, n-1} & = & \frac{
S_{x_{n-1}} ( \{ k^{new}_{x_{n-1},\bullet}  \}) \; 
S_{x_0}       ( \{ k^{new}_{x_0,\bullet}  \}) } {A} \;
\frac{ L_{x_{n-1},\nu_{n-1}}(k_{x_{n-1},\nu_{n-1}}^{trial})}{L_{x_{n-1},\nu_{n-1}}(k_{x_{n-1},\nu_{n-1}})} \; .
\label{rhoT}
\end{eqnarray}
Here we use the notation $S_{x_j} ( \{ k^{new}_{x_j,\bullet} \})$ to indicate that in the set of link variables
$\{k_{x_j,\bullet} \}$, which are the arguments of the site weights 
$S_{x_j} (\{k_{x_j,\bullet} \})$, one inserts the values the fluxes have after the worm has passed through $x_j$. More explicitly:
\begin{equation}
\{ k^{new}_{x_j,\bullet} \} \; \equiv \; \{\, k_{x_j,\bullet} \} \;\; \mbox{with} \; \;
 k_{x_{j},-\nu_{j-1}} \leftarrow k_{x_{j},-\nu_{j-1}} + 
\,\mbox{sign}(\nu_{j-1}) \Delta \; \; \mbox{and} \; \; 
k_{x_{j},\nu_{j}} \leftarrow k^{trial}_{x_{j},\nu_{j}} \; ,
\end{equation}
i.e., the flux in the previous link $(x_{j-1},\hat{\nu}_{j-1}) = (x_j,-\hat{\nu}_{j-1})$ of the worm has been replaced by 
the already accepted new flux $k_{x_{j},-\nu_{j-1}} + \,\mbox{sign}(\nu_{j-1}) \Delta$ and for the current step we use 
$k_{x_j,\nu_j}^{trial} = k_{x_j,\nu_j} +  \,\mbox{sign}(\nu_{j}) \Delta$. 

The real and positive 
amplitude $A$ in (\ref{rhoS}) and (\ref{rhoT}) is a free parameter of the algorithm and we will show that detailed balance
holds for arbitrary positive values of $A$. It can be used to control the acceptance probability of the first step 
and the terminating probability of the last step: In a region of couplings where the average
value of the site weights $S_{x_j} (\{k_{x_j,\bullet} \})$ is very high, the acceptance probability for the initial step 
would be very small, and increasing the amplitude $A$ can be used for boosting the starting probability of the worm. 

On the other hand, for couplings where the average value of the site weights $S_{x_j} (\{k_{x_j,\bullet} \})$ is very small, 
the acceptance probability for the terminating step would also be small, giving rise to worms that propagate for many steps
(not necessarily changing many fluxes, since a worm can annihilate its own changes via dangling dead ends). This implies 
that adjusting $A$ gives us control over the average length of worm and thus allows one to tune the efficiency of the 
algorithm. 

Before we come to proving detailed balance for the worm defined by the Metropolis ratios, let us comment a little on the choices
(\ref{rhoS}) -- (\ref{rhoT}). For a system without site weights, i.e., $S_{x} (\{k_{x,\bullet} \}) = 1 \, \forall x$, our worm algorithm reduces
to the standard worm algorithm \cite{worm}, which only depends on the ratios of the link weights $L_{x,\nu}(k_{x,\nu})$. In these ratios
the weights are also balanced, such that the amplitude factor $A$ is not necessarily needed, but still may be introduced to adjust 
the average length of the worm. 

In the general wordline model (\ref{Zflux}) also the site weights $S_{x} (\{k_{x,\bullet} \})$ need to be taken into account for the 
Metropolis ratios. However, here we have the situation that when the worm proposes the new flux 
$k_{x_j,\nu_j}^{trial}  = k_{x_j,\nu_j} + \, \mbox{sign}(\nu_j) \, \Delta$ for the current link $(x_j,\nu_j)$, this proposal flux
enters the arguments at both endpoints of that link, i.e., at $x_j$ and $x_{j+1} = x_j + \hat{\nu}_j$. The problem is that the final
value of $S_{x_{j+1}} (\{k^{new}_{x_{j+1},\bullet} \})$ is not yet known\footnote{The special form of the $k_{x,\nu}$-dependence
in the RBG admits an alternative strategy based on the insertion of an initial source-sink 
pair and a different propagation rule for the defect
\cite{Rindlisbacher}.}, since it also depends on the decision the worm makes 
in the subsequent step, i.e., on the choice of the link $(x_{j+1},\nu_{j+1})$. Thus in the Metropolis ratio $\rho_{x_j,\nu_j}^{\,R}$
for the link $(x_j,\nu_j)$ we can take into account the final site weight $S_{x_j} ( \{ k^{new}_{x_j,\bullet} \})$ 
only for the starting point $x_j$ of that link. In the Metropolis ratio $\rho^{\, R}_{x_j,\nu_j}$ in (\ref{rhoR}) we form the ratio of that 
final site weight $S_{x_j} ( \{ k^{new}_{x_j,\bullet}  \})$ at $x_j$ with the old site weight 
$S_{x_{j+1}} ( \{ k_{x_{j+1},\bullet} \})$ at $x_{j+1}$ in the denominator.

The situation is even more involved for the starting Metropolis ratio $\rho^S$ defined in (\ref{rhoS}): 
There we also do not know the final site weight
for the starting site $x_0$, as its value will be determined only in the last step of the worm, depending on the direction $\nu_{n-1}$
in which the closing site $x_n = x_0$ is approached and the flux $k_{x_{n-1},\nu_{n-1}}$ of the last link in the worm.

\section{Proof of detailed balance}

We are now ready to prove the detailed balance condition (\ref{detailedbalance}) with the weight $W(C_k)$
of the flux configurations $C_k$ given in (\ref{PC}) and the worm algorithm defined by the pseudocode in the previous section with the
Metropolis ratios (\ref{rhoS}) -- (\ref{rhoT}) for starting, running and terminating the worm. We have already stressed that there are 
infinitely many worms $w$ that lead from a configuration $C_k$ to the new configuration $\widetilde{C}_k$ which differs
by the flux along some loop $L$. 
The total transition probability $P(C_k \rightarrow \widetilde{C}_k)$ is a 
sum over the individual contributions $P_w (C_k \rightarrow \widetilde{C}_k)$ from the worms $w$,  \begin{equation}
P(C_k \rightarrow \widetilde{C}_k) \; = \; \sum_{\{w\}} P_w (C_k \rightarrow \widetilde{C}_k) \; .
\end{equation}
In the same way we have infinitely many reverse worms $w^{\star}$ that lead from $\widetilde{C}_k$ to $C_k$ with individual
contributions $P_{w^\star} (\widetilde{C}_k \rightarrow  C_k)$ to the reverse total transition probability 
$P(\widetilde{C}_k \rightarrow  C_k)$. Thus, when taking into account all worms connecting 
$C_k$ and $\widetilde{C}_k$, the detailed balance condition (\ref{detailedbalance}) assumes the form
\begin{equation}
\frac{\sum_{\{w\}} \,\, P_w \, (C_k \rightarrow \widetilde{C}_k)}{
\sum_{\{w^\star\}} P_{w^\star} (\widetilde{C}_k \rightarrow C_k)} \; = \; \frac{W(\widetilde{C}_k)}{W(C_k)} \; .
\label{detailedbalance2}
\end{equation}
The key observation for finding a sufficient condition to solve this equation is the fact that the set $\{w\}$ of worms from 
$C_k$ to $\widetilde{C}_k$ and the set  $\{w^{\star}\}$ of worms from $\widetilde{C}_k$ to $C_k$ allow for 
one-to-one mappings, such that to every worm $w$ in $\{w\}$ we can identify a worm $w^{-1} \in \{w^\star \}$, which we
refer to as inverse worm. For a given worm $w$, we define the inverse worm $w^{-1}$ via
\begin{equation}
w \; = \; (  x_0, \nu_0 , \Delta, \nu_1, \, ... \, \nu_{n-1}) \quad \rightarrow \quad 
w^{-1} \; \equiv \; ( x_0,-\nu_{n-1},  \Delta, -\nu_{n-2}, \, ... \, - \nu_1, -\nu_0) \; ,
\label{inverseworm}
\end{equation}
i.e, the inverse worm $w^{-1}$ starts at the same site $x_0$ and runs backwards through the steps of $w$, using the same flux
increment $\Delta$. It is obvious, that $w^{-1}$ inverts the steps of $w$ and changes the configuration $\widetilde{C}_k$
back into $C_k$. 

Using the fact that our definition of $w^{-1}$ induces a one-to-one mapping between $\{w^{\star}\}$ and $\{w\}$, we can rewrite the 
detailed balance condition (\ref{detailedbalance2}) into the form,
\begin{equation}
\sum_{\{w\}}  P_w (C_k \rightarrow \widetilde{C}_k) \; = \; \frac{W(\widetilde{C}_k)}{W(C_k)} \;
\sum_{\{w^\star\}} P_{w^\star} (\widetilde{C}_k \rightarrow C_k) \; = \; \frac{W(\widetilde{C}_k)}{W(C_k)} \; 
\sum_{\{w\}} P_{w^{-1}} (\widetilde{C}_k \rightarrow C_k) \; .
\label{detailedbalance3}
\end{equation}
An obvious sufficient solution to this equation is the condition
\begin{equation}
\frac{P_w (C_k \rightarrow \widetilde{C}_k)}{P_{w^{-1}} (\widetilde{C}_k \rightarrow C_k)} \; = \; 
\frac{W(\widetilde{C}_k)}{W(C_k)} \qquad \forall \, w \; ,
\label{ratioworm}
\end{equation}
which demands that the ratio of the transition probability $P_w (C_k \rightarrow \widetilde{C}_k)$ of $w$ and the transition 
probability $P_{w^{-1}} (\widetilde{C}_k \rightarrow C_k)$ of the inverse worm $w^{-1}$ equals the ratio 
$W(\widetilde{C}_k)/W(C_k)$ of weights of the two configurations  $\widetilde{C}_k$ and $C_k$. 

Note that one could be tempted to choose different definitions for the inverse of a worm $w$, e.g., the worm
$w^{\neg} \equiv ( x_0, \nu_0, -\Delta, \nu_1, \, ... \, \nu_{n-1})$ 
which starts at the same site $x_0$, runs through the same links in the same
orientation as $w$, but 
with the negative flux increment $-\Delta$.  However, $w^{-1}$ as defined in (\ref{inverseworm}) is the natural choice if we take 
into account how the cancellation proceeds: The inverse worm $w^{-1}$ starts with the closing link of $w$ and step by step 
cancels the changes of $w$. This implies that for every site on the contour of the worm 
the heads of $w$ and of $w^{-1}$ see the same 
configuration of fluxes and thus have the same options for steps with the same weights. As we will see, this property is essential for 
the correct normalization of the probabilities of the individual steps of the worm. 
For the alternative worm $w^{\neg}$ this is not the case, and for every site
on the worm contour $w$ and $w^{\neg}$ see a different configuration of flux. Thus $w$ and $w^{\neg}$ have different 
probabilities for their choices, and the cancellation of the normalization factors, which is an essential step in the proof of detailed 
balance below, is no longer possible.

The transition probability $P_w (C_k \rightarrow \widetilde{C}_k)$ of a worm $w$ is given by a product over the 
probabilities $P^S_{x_0,\nu_0}$, $P^R_{x_j,\nu_j}$ and $P^T_{x_{n-1},\nu_{n-1}}$ for its steps, 
\begin{equation}
P_w (C_k \rightarrow \widetilde{C}_k) \; = \; P^S_{x_0,\nu_0} \; \left( \prod_{j=1}^{n-2} P^R_{x_j,\nu_j} \right) \; \, 
P^T_{x_{n-1},\nu_{n-1}} \; .
\label{stepsofworm}
\end{equation}
Using this form and the identification of the inverse worm $w^{-1}$ given in (\ref{inverseworm}), Equation (\ref{ratioworm}) turns into 
\begin{equation}
\frac{P_w (C_k \rightarrow \widetilde{C}_k)}{P_{w^{-1}} (\widetilde{C}_k \rightarrow C_k)} \; = \; 
\frac{ P^S_{x_0,\nu_0} \; \left( \prod_{j=1}^{n-2} P^R_{x_j,\nu_j} \right) \; \, P^T_{x_{n-1},\nu_{n-1}} }
{\tilde P^S_{x_0,-\nu_{n-1}} \; \left( \prod_{j={n-1}}^{2} \tilde P^R_{x_j,-\nu_{j-1}} \right) \; \, \tilde P^{\,T}_{x_{1},-\nu_{0}} }  
\; = \; \frac{W(\widetilde{C}_k)}{W(C_k)} \; .
\label{ratioworm2}
\end{equation}
Here we use the notation $\tilde P^S_{x_0,-\nu_{n-1}}$, $\tilde P^R_{x_j,-\nu_{j-1}}$ and $\tilde P^{\,T}_{x_{1},-\nu_{0}}$ 
to indicate that the probabilities of the inverse worm $w^{-1}$, appearing in the denominator of (\ref{ratioworm2}), start from the 
configuration $\widetilde{C}_k$ which the worm $w^{-1}$ takes back into the original configuration $C_k$. 

Let us now come to discussing the probabilities $P^S_{x_0,\nu_0}$, $P^R_{x_j,\nu_j}$ and $P^T_{x_{n-1},\nu_{n-1}}$ for
the individual steps of the worms. It is important to understand the difference between the 
Metropolis ratios $\rho^{\, S}$, $\rho^{\, R}$ and $\rho^{\, T}$ given in 
(\ref{rhoS}) -- (\ref{rhoT}) and the step probabilities $P^S_{x_0,\nu_0}$, $P^R_{x_j,\nu_j}$ and $P^T_{x_{n-1},\nu_{n-1}}$ which
constitute the probability $P_w(C_k \rightarrow \widetilde{C}_k)$ in Eq.~(\ref{stepsofworm}). 
The true, normalized probability of the worm to go in a new direction is not given by the Metropolis acceptance probability 
min$\{\rho,1\}$ alone. One has to take into account the "try until accepted" feature for the intermediate 
and terminating steps in the worm algorithm\footnote{As we show here, the ''try until accepted" feature gives rise to a correct algorithm.
However, a correct algorithm is not necessarily efficient: in principle the worm could encounter some configurations where it
gets stuck at a site $x$, since the proposed steps out of $x$ are almost always rejected. A possible way out would be to replace 
the "try until accepted" steps by a heatbath step.}, and for these steps the Metropolis acceptance
probability min$\{\rho,1\}$ has to be normalized by a sum over the Metropolis acceptance probabilities of all possible steps. 
Consequently the probabilities for the steps in the worm $w$ are given by
\begin{equation}
P^S_{x_0,\nu_0} \; = \; \frac{ \mbox{min} \, \Big\{ \rho^{S}_{x_0,\nu_0} , 1 \Big\} }{2dV} \quad , \quad 
P^R_{x_j,\nu_j} \; = \; \frac{ \mbox{min} \, \Big\{  \rho^{R}_{x_j,\nu_j}, 1 \Big\} }{ N_{x_j} } \quad , \quad 
P^T_{x_j,\nu_j} \; = \; \frac{ \mbox{min} \, \Big\{  \rho^{T}_{x_j,\nu_j}, 1  \Big\} }{ N_{x_j} } \; ,
\label{eq:Prob}
\end{equation}
where the normalization $N_{x_j}$ for the probabilities of the steps for running and terminating the worm is given by 
\begin{equation}
N_{x_i} \, = \; \sum_{\sigma = \pm 1}^{\pm d} \Big[ \mbox{min} \, \Big\{  \rho^{R}_{x_i,\sigma}, 1 \Big\} [ 1 -  \delta_{x_i+\hat{\sigma},x_0} ]  \; + \; \mbox{min} \, \Big\{  \rho^{T}_{x_i,\sigma}, 1 \Big\} \delta_{x_i+\hat{\sigma},x_0} \Big] ~.
\end{equation}
As discussed, this takes into account the "try until accepted" feature of the worm in the given configuration of fluxes. 
The probability for the starting step of the worm has no "try until accepted" feature, such that it is given by the Metropolis
acceptance probability $\mbox{min} \, \Big\{ \rho^{S}_{x_0,\nu_0} , 1 \Big\}$ normalized with the probability $1/2dV$ 
for selecting the starting link and its orientation. We furthermore stress that using the probabilities defined in (\ref{eq:Prob}) it is possible to demonstrate the normalization condition for the full transition probabilities $\sum_{\widetilde{C}_k} P (C_k \rightarrow \widetilde{C}_k) = 1$.

Now we also see why the particular choice for inverse worm $w^{-1}$ in Eq.~(\ref{inverseworm}) 
is fundamental for showing detailed balance. 
For this choice of $w^{-1}$ one has exactly the same normalization factors as in $w$, and all factors $1/N_{x_j}$ cancel in 
(\ref{ratioworm2}).
A different choice of $w^{-1}$, e.g., as the worm $w^{\neg}$ that runs through the steps
of $w$ in the same orientation but with $-\Delta$, does not give rise to this cancellation. 

Having established the cancellation of the normalization factors in (\ref{ratioworm2}), we are left with showing
\begin{equation}
\frac{P_w (C_k \rightarrow \widetilde{C}_k)}{P_{w^{-1}} (\widetilde{C}_k \rightarrow C_k)} \; = \; 
\frac{ \, \mbox{min} \{\rho^S_{x_0,\nu_0} , 1 \} }{ \, \mbox{min} \{ \tilde \rho^{\,T}_{x_{1},-\nu_{0}} , 1 \} } \!
\left( \prod_{j=1}^{n-2} 
\frac{\, \mbox{min} \{ \rho^R_{x_j,\nu_j}, 1 \} } { \, \mbox{min} \{ \tilde \rho^R_{x_{j+1},-\nu_{j}},1\} } \! \right) \! 
\frac{  \, \mbox{min} \{\rho^T_{x_{n-1},\nu_{n-1}} , 1 \} } { \, \mbox{min} \{ \tilde\rho^S_{x_0,-\nu_{n-1}}, 1 \} }
\; = \; \frac{W(\widetilde{C}_k)}{W(C_k)} .
\label{ratioworm3}
\end{equation}
Here we have already reordered the terms of the inverse worm $w^{-1}$ in the denominators and paired them in ratios with 
the matching links of the forward worm in the numerators.
Inspecting the Metropolis ratios (\ref{rhoS}) -- (\ref{rhoT}) for a worm $w$ and the corresponding inverse worm $w^{-1}$ defined
in (\ref{inverseworm}), one easily shows the following properties,
\begin{equation}
\tilde \rho^{\,T}_{x_{1},-\nu_{0}} \; = \; \frac{1}{\rho^S_{x_0,\nu_0}} \; , \; \; \quad
\tilde \rho^R_{x_{j+1},-\nu_{j}} \; = \; \frac{1}{\rho^R_{x_j,\nu_j}} \; , \; \; \quad
\tilde\rho^S_{x_0,-\nu_{n-1}} \; = \; \frac{1}{\rho^T_{x_{n-1},\nu_{n-1}}} \; .
\end{equation}
Finally we make use of the identity $\mbox{min} \{ \rho, 1\} / \mbox{min} \{ \rho^{-1}, 1\} = \rho$ for $\rho > 0$  and find
\begin{equation}
\frac{P_w (C_k \rightarrow \widetilde{C}_k)}{P_{w^{-1}} (\widetilde{C}_k \rightarrow C_k)} \; = \; 
\rho^S_{x_0,\nu_0} 
\left( \prod_{j=1}^{n-2} \rho^R_{x_j,\nu_j} \right) \rho^T_{x_{n-1},\nu_{n-1}} \; = \; \frac{W(\widetilde{C}_k)}{W(C_k)} \; .
\label{ratioworm4}
\end{equation}
Using (\ref{rhoS}) -- (\ref{rhoT}) for the Metropolis ratios and multiplying them as in the second term of (\ref{ratioworm4}) 
one finds that the result for this product is indeed the rhs.\  $W(\widetilde{C}_k)/W(C_k)$. Links in dangling dead ends appear in 
forward, as well as in backward direction in the product and it is easy to see that these factors cancel, and do not contribute to 
$W(\widetilde{C}_k)/W(C_k)$. In the same way it is easy to see that also the amplitude factor $A$ cancels and can be chosen 
freely to optimize the starting and closing probabilities and thus the average length of the worms. 
This concludes the proof of detailed balance. 
 
\section{Exploring the role of the amplitude parameter $A$}

\begin{figure}[p!]
\begin{center}
\vspace*{-10mm}
\includegraphics[width=7.8cm,clip]{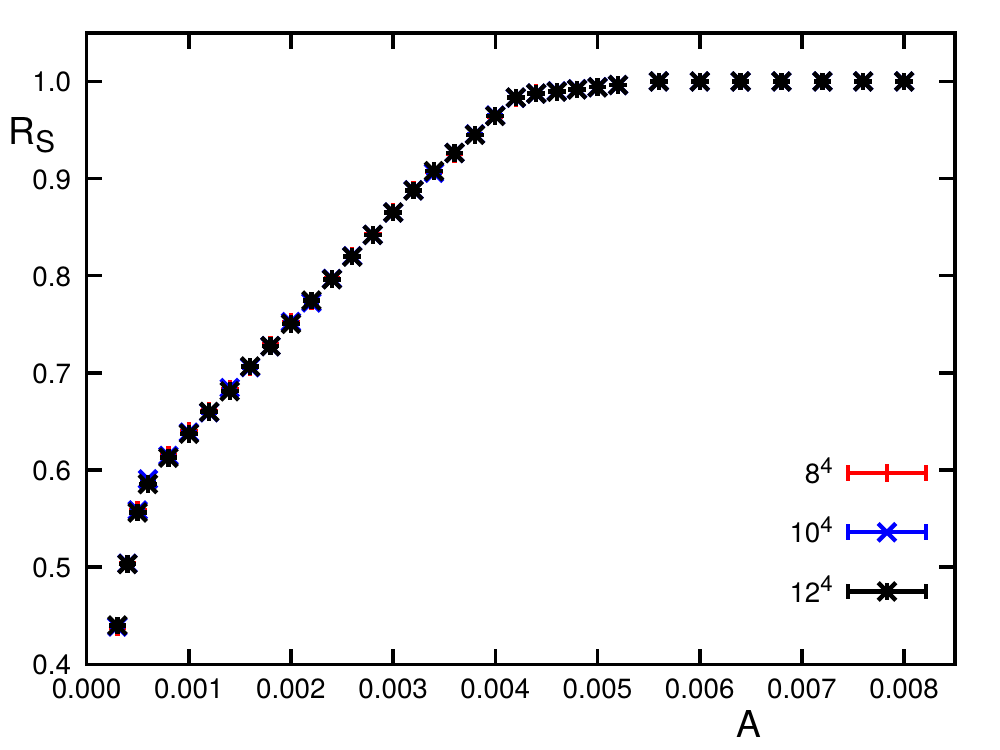}
\hspace{2mm}
\includegraphics[width=7.85cm,clip]{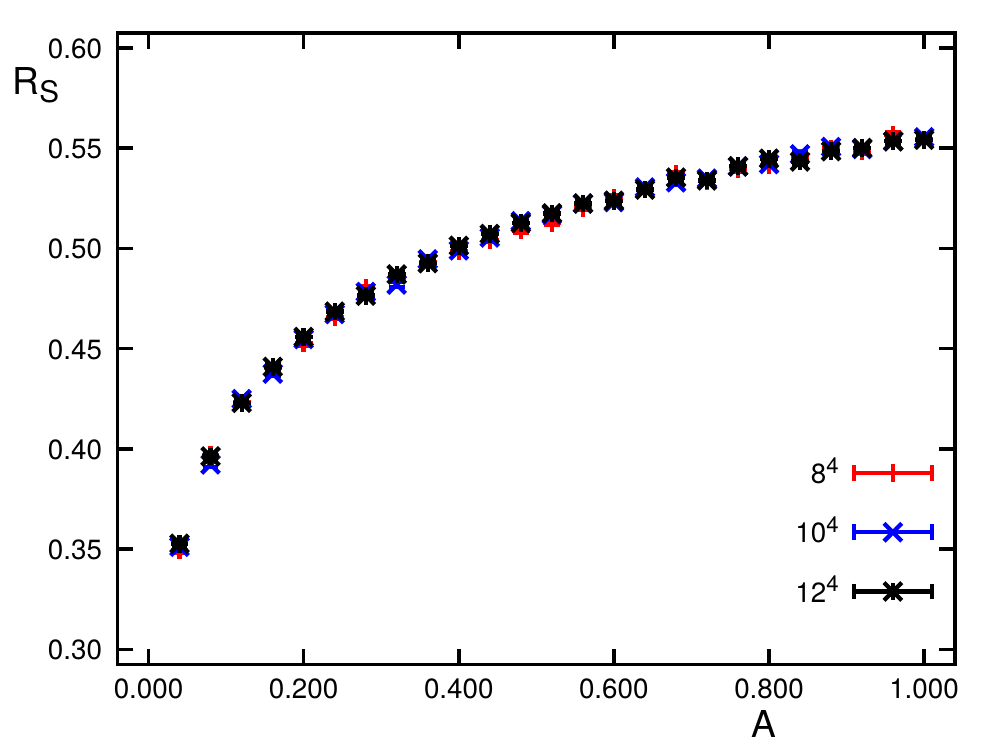}
\end{center}
\vspace{-6mm}
\caption{Starting probability $R_S$ as a function of the amplitude $A$. We show the 
results for coupling set I (lhs.\ plot) and coupling set II (rhs.) and compare different volumes.}
\label{starting}

\begin{center}
\includegraphics[width=7.8cm,clip]{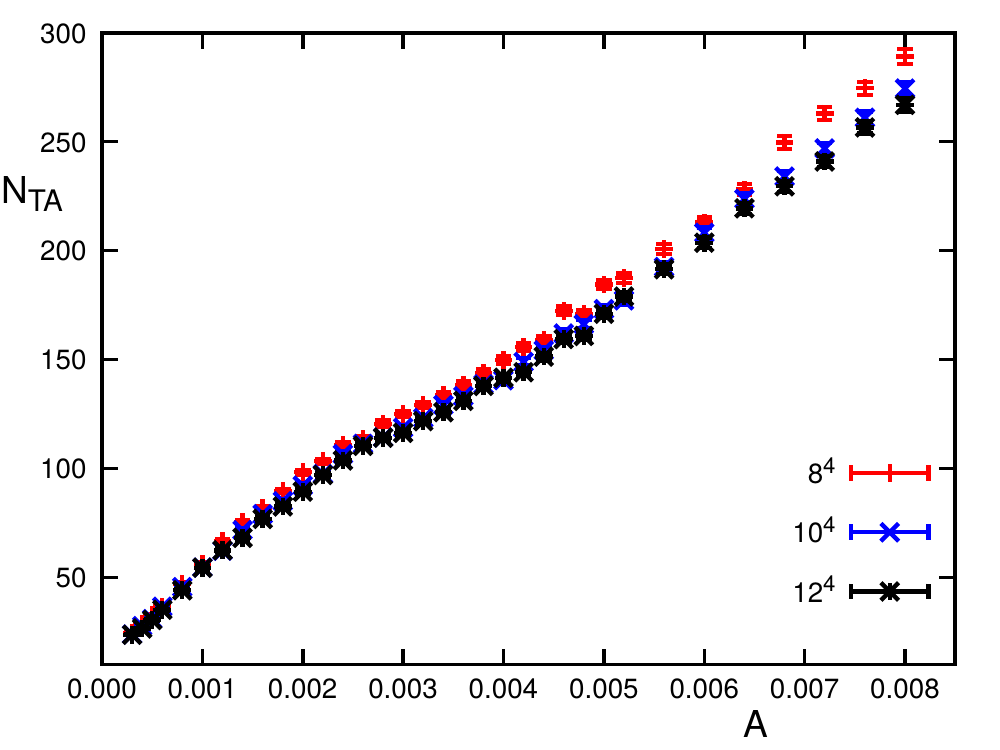}
\hspace{2mm}
\includegraphics[width=7.8cm,clip]{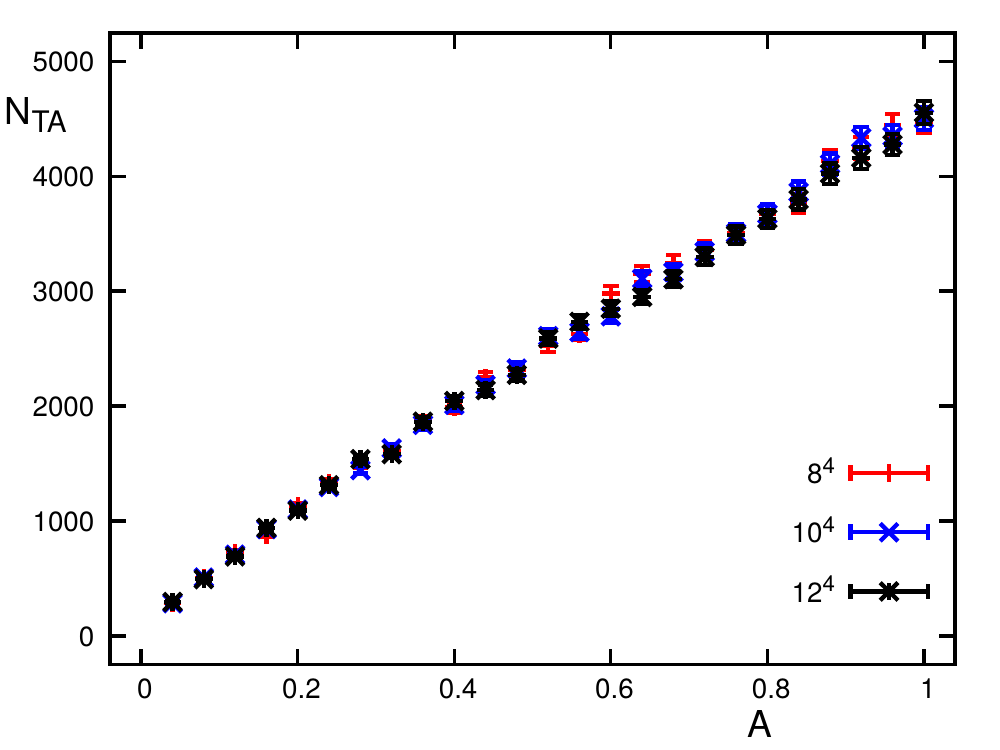}
\end{center}
\vspace{-6mm}
\caption{Average number of termination attempts $N_{TA}$ as a function of the amplitude $A$. We show the 
results for coupling set I (lhs.\ plot) and coupling set II (rhs.) and compare different volumes.}
\label{terminating}

\begin{center}
\includegraphics[width=7.8cm,clip]{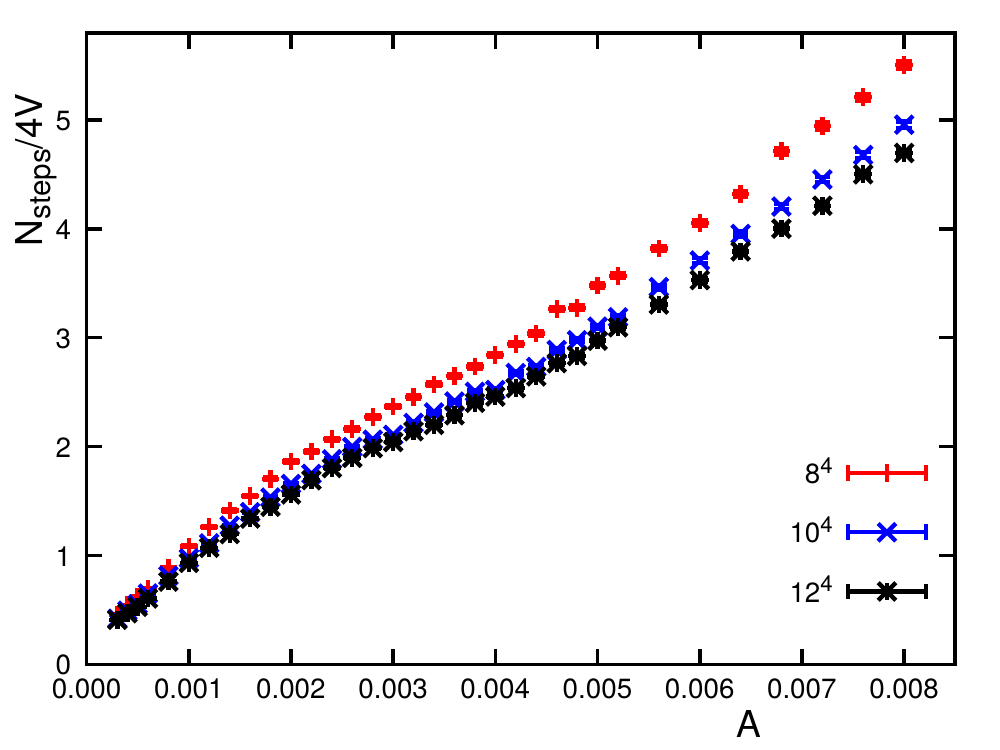}
\hspace{2mm}
\includegraphics[width=7.86cm,clip]{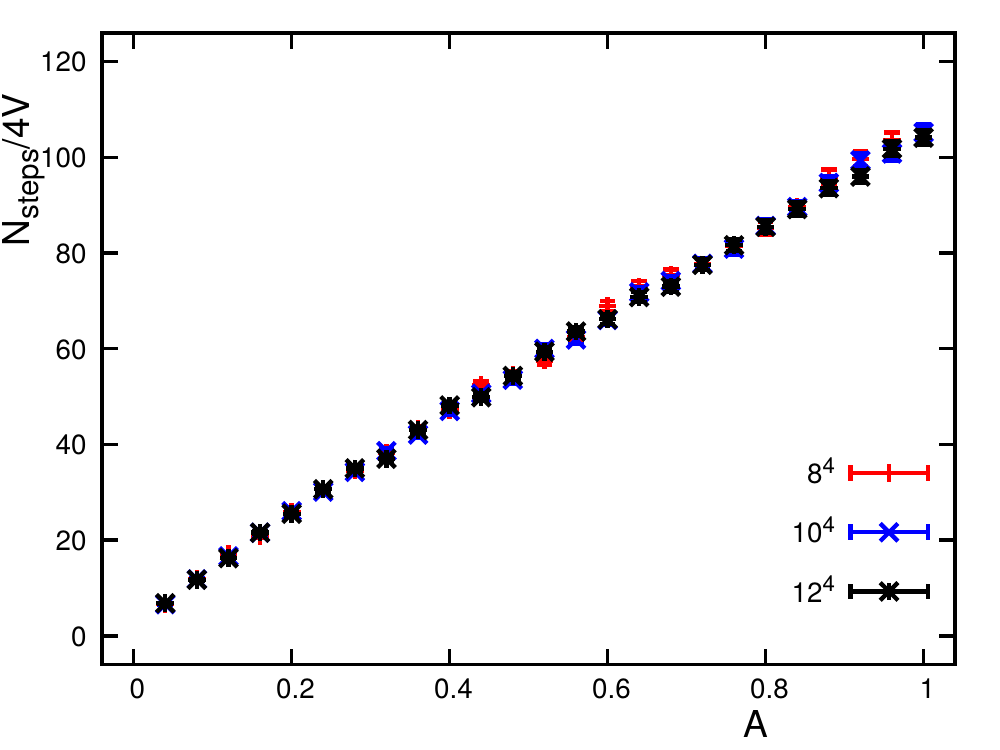}
\end{center}
\vspace{-6mm}
\caption{Number of steps per worm $N_{step}$ normalized with $4V$, the number of lattice links as a function of the amplitude $A$.
We show the results for coupling set I (lhs.\ plot) and coupling set II (rhs.) and compare different volumes.}
\label{steps}

\end{figure}

\begin{figure}[p!]
\begin{center}
\vspace*{-10mm}
\includegraphics[width=7.8cm,clip]{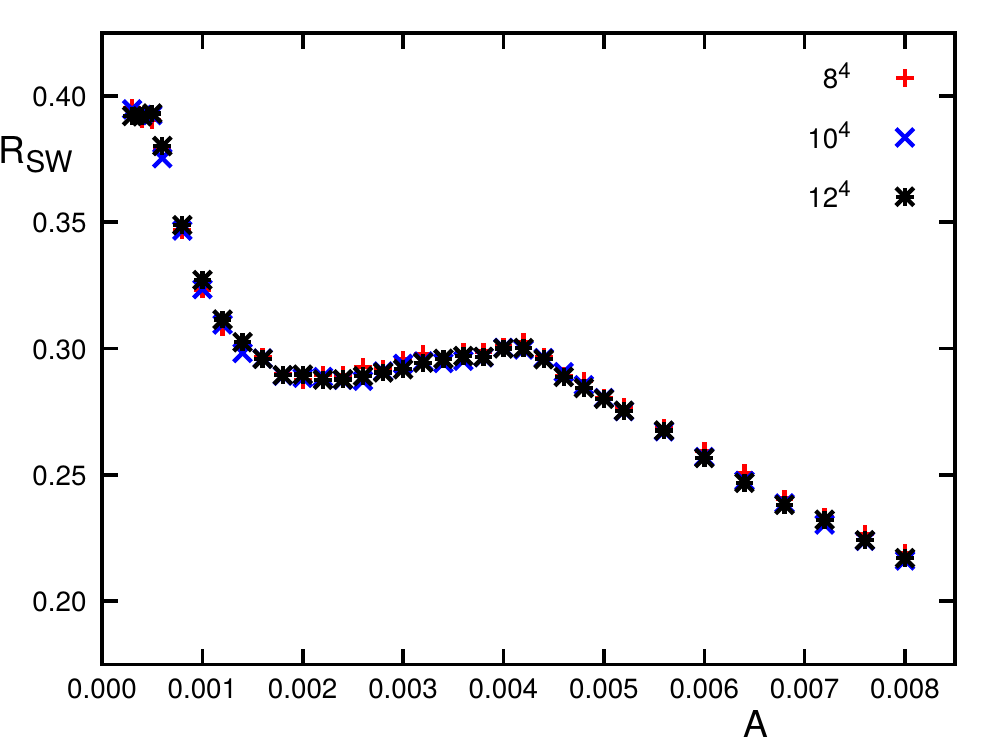}
\hspace{2mm}
\includegraphics[width=7.83cm,clip]{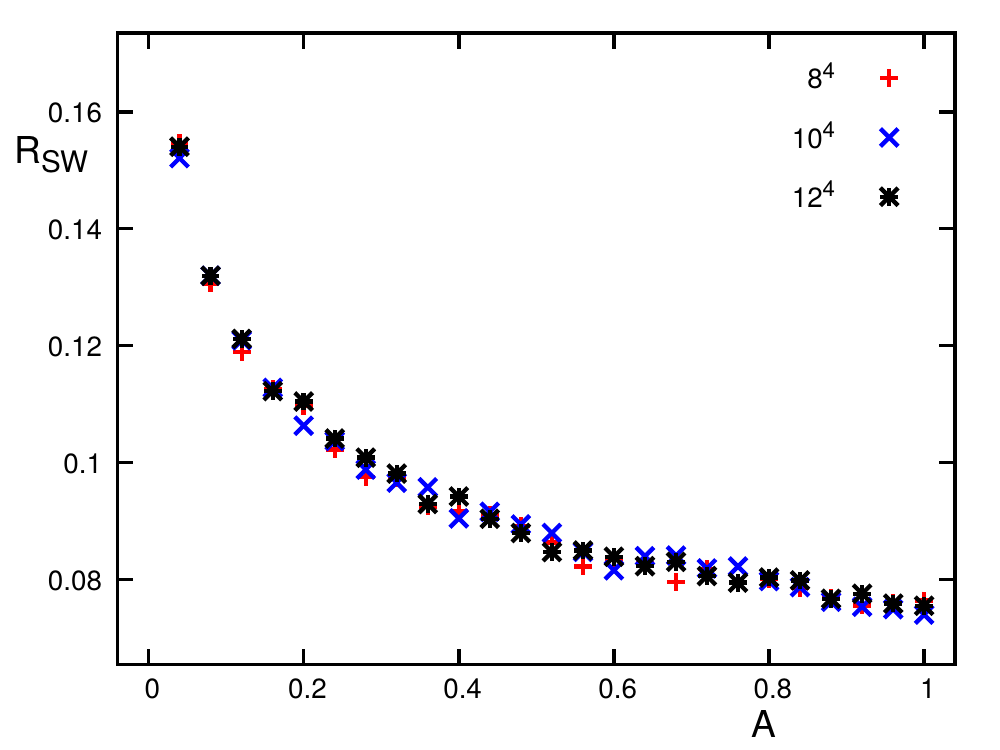}
\end{center}
\vspace{-6mm}
\caption{Ratio of sterile worms $R_{SW}$ as a function of $A$. We show the results for coupling set I (lhs.\ plot) 
and coupling set II (rhs.) and compare different volumes.}
\label{zero}

\begin{center}
\includegraphics[width=7.8cm,clip]{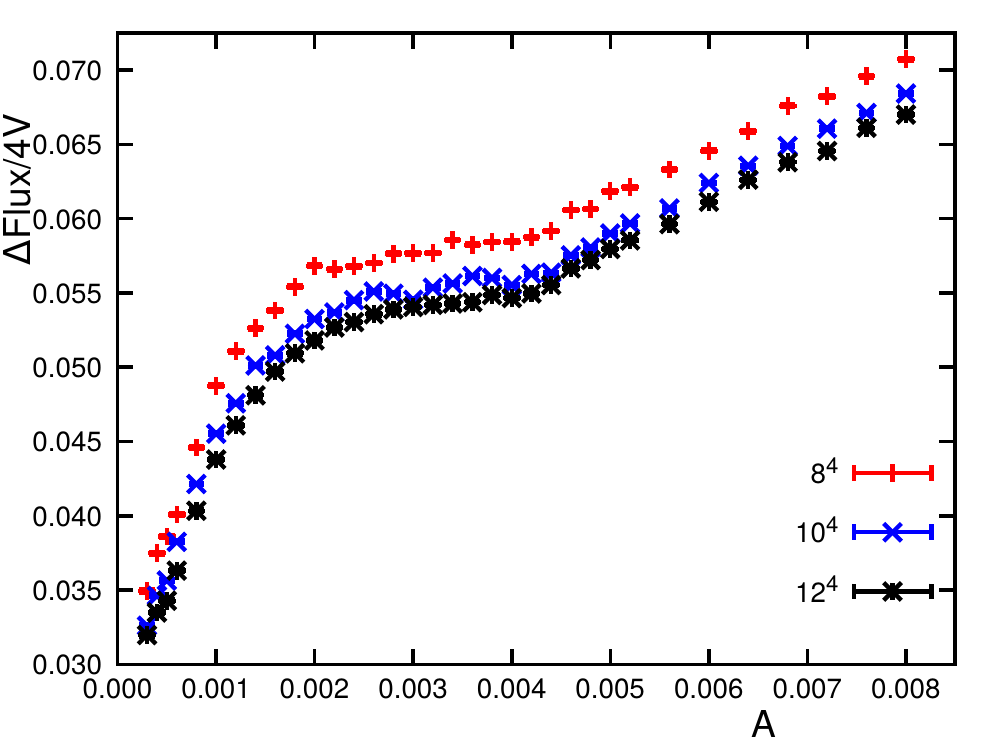}
\includegraphics[width=7.85cm,clip]{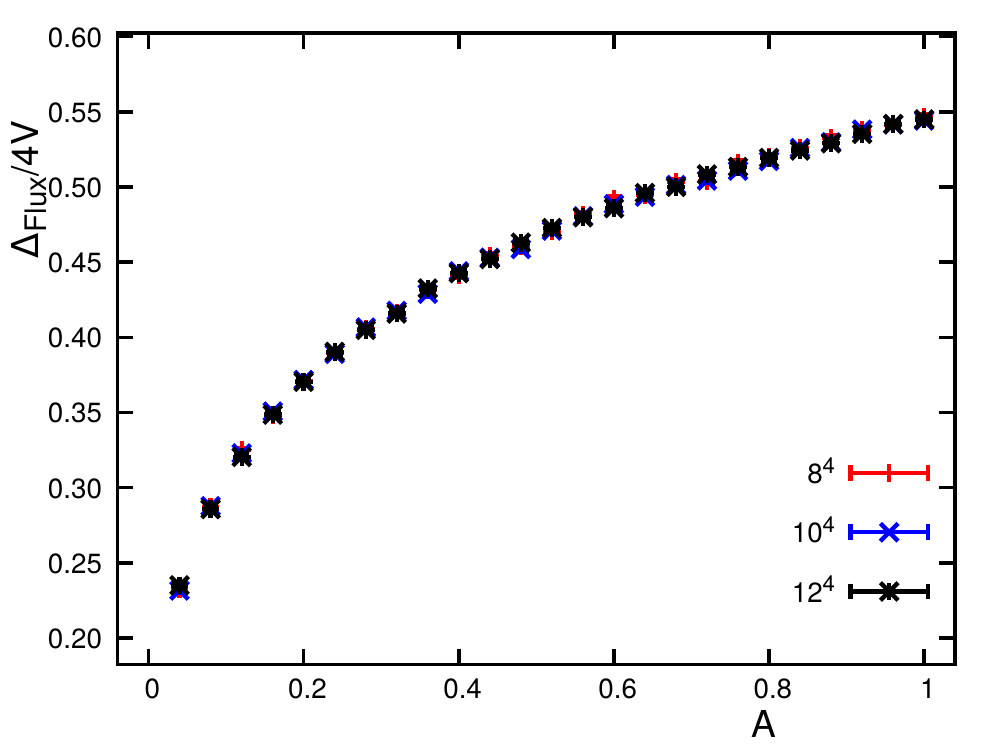}
\end{center}
\vspace{-6mm}
\caption{Average change of flux $\Delta_{flux}$ normalized by the number of lattice links as a function of $A$. 
We show the results for coupling set I (lhs.\ plot) and set II (rhs.), comparing different volumes.}
\label{length}

\begin{center}
\includegraphics[width=7.8cm,clip]{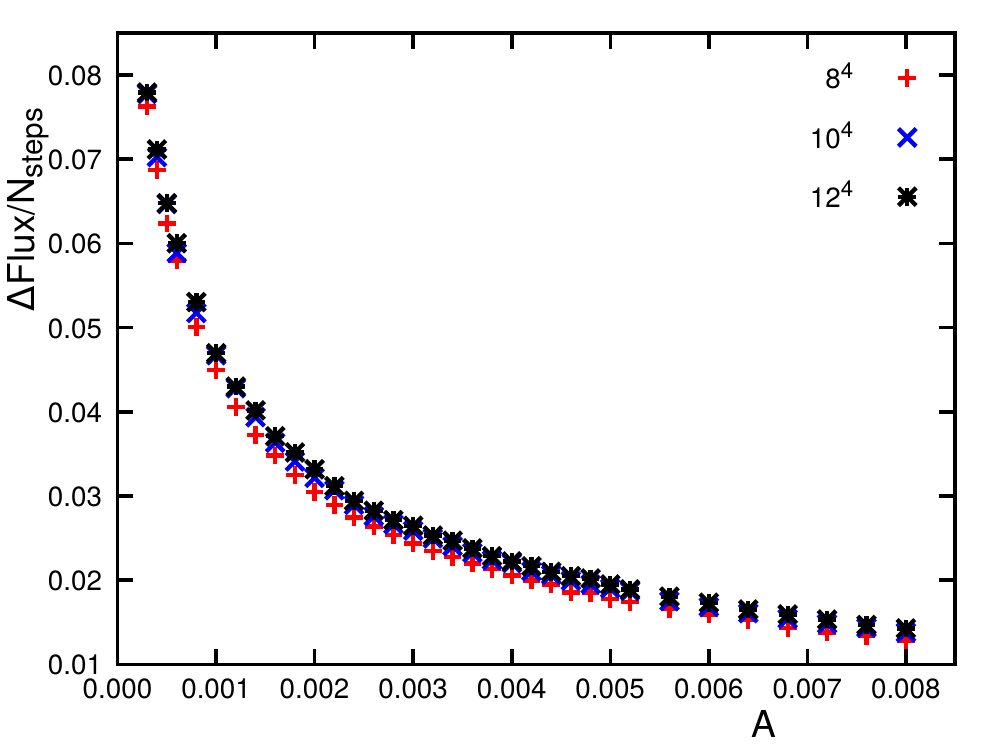}
\includegraphics[width=7.85cm,clip]{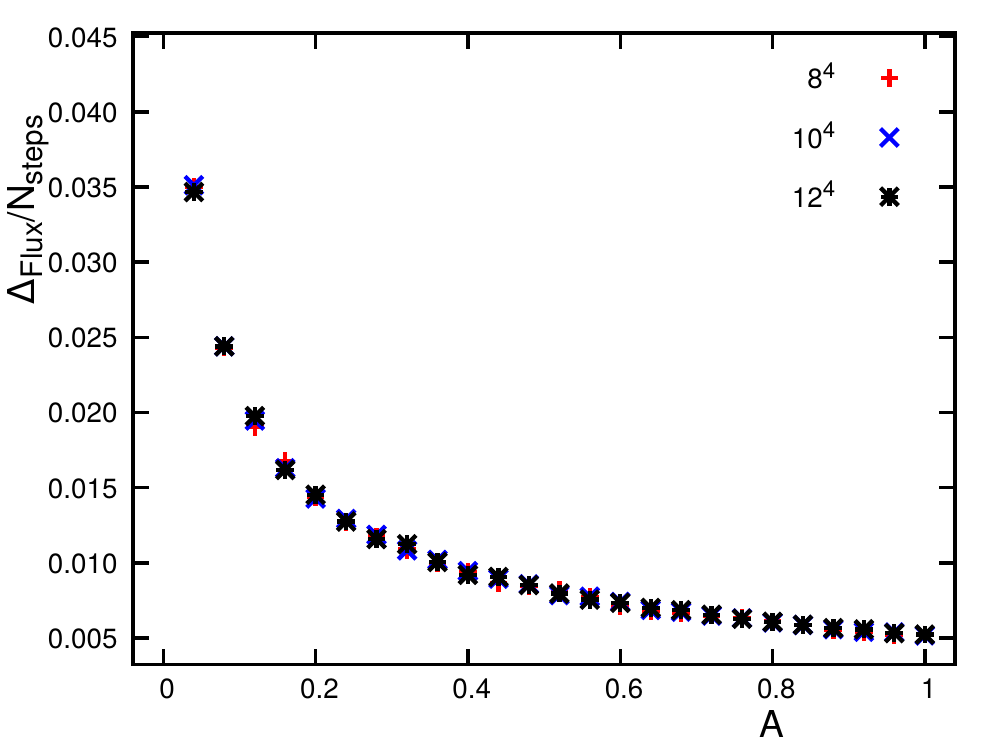}
\end{center}
\vspace{-6mm}
\caption{Average change of flux $\Delta_{flux}$ normalized by the number of steps of the worm as a function of $A$.
We show the results for coupling set I (lhs.\ plot) and set II (rhs.), comparing different volumes.}
\label{change}

\end{figure}

Let us now analyze the role of the amplitude parameter $A$. We have seen that detailed balance holds independently of
the actual value of $A$. However, the amplitude enters in the Metropolis ratios (\ref{rhoS}) and 
(\ref{rhoT}) for starting and for terminating the worms. It is obvious, that increasing $A$ to sufficiently large values will
increase the acceptance probability min$\{ \rho^S, 1\}$ for the starting step. On the other hand, increasing $A$
will lower the probability min$\{ \rho^T, 1\}$ for accepting the terminating step. Thus we expect that large $A$ increases 
the length of the worms, since the terminating step is accepted less often and more termination attempts are needed to
finish the worm.

This qualitative picture can be checked with the following quantities: The starting ratio $R_S$, the number of 
termination attempts $N_{TA}$, the number of steps $N_{steps}$, and the ratio of sterile worms $R_{SW}$. 
They are defined as follows:
\begin{eqnarray}
R_S & = & \frac{\mbox{number of started worms}}{\mbox{number of starting attempts}} \;\; ,
\label{observables} \\
& & \nonumber \\
N_{TA} & = & \mbox{number of attempts to terminate the worm until the worm closes} \;\; ,
\nonumber \\
& & \nonumber \\
N_{steps} & = & \mbox{number of accepted steps in the worm \quad (this is the number $n$ in (\ref{wormnotation}))}  \;\; ,
\nonumber \\
& & \nonumber \\
R_{SW} & = &  \frac{\mbox{number of worms that do not change the configuration}}{\mbox{number of  all started worms}} \;\; .
\nonumber \\
& & \nonumber 
\end{eqnarray}

In order to analyze the role of the amplitude parameter $A$ in our algorithm we implemented the worm for the relativistic
Bose gas as defined in (\ref{action}) using the dual representation (\ref{bosedual}) and a conventional local Monte Carlo 
update for the auxiliary variables $l_{x,\nu}$. Here sweeps for the $l_{x,\nu}$ are alternated with worms, but it is also
possible to interleaf worm steps and updates of the auxiliary variables \cite{Rindlisbacher}. 
The correctness of the implementation of the 
worm algorithm was tested against analytical results for the free case of $\lambda = 0$ and against the numerical results of a
standard Monte Carlo simulation in the conventional representation at $\mu = 0$. For the analysis of the algorithmic observables 
(\ref{observables}) that illustrate the role of the amplitude we use two set of couplings:

\vskip3mm

Coupling set I: $\;\,\,\eta = 7.44$, $\lambda  = 1.0$,  $\mu = 0.275 \; \; , \; \; \; \; $ 
Coupling set II: $\;\eta = 6.00$, $\lambda  = 0.4$,  $\mu = 0.275 \; \; .$

\vskip3mm

\noindent
The statistics is $5 \times 10^5$ worms after $5 \times 10^4$ worms for equilibration for coupling set I, and 
$10^5$ worms after $10^4$ worms for equilibration for coupling set II. The algorithmic observables are computed
for three different lattice sizes $8^4$, $10^4$ and $12^4$ to assess possible finite size effects on the worm. 

We stress at this point that this numerical analysis is not meant as a detailed study of autocorrelation properties 
(which also would imply a more systematic scan of the coupling space). The simulations presented here focus on
illustrating in a specific model at two different coupling sets
that the amplitude parameter $A$ indeed can be used to vary and optimize
properties of the worm algorithm.

The quantities that directly analyze the starting and terminating probabilities, i.e., the starting ratio $R_S$ and the number of
terminating attempts $N_{TA}$ are shown in Fig.~\ref{starting} and in Fig.~\ref{terminating}. 
Both quantities are plotted as a function of $A$ and we show the results for coupling sets I and II for the three different volumes.  
The starting ratio $R_S$  (Fig.~\ref{starting}), as well as the number of termination attempts $N_{TA}$ (Fig.~\ref{terminating}) 
show the expected dependence on $A$: $R_S$ is an increasing function 
of $A$ and for coupling set I (lhs.\ plot in Fig.~\ref{starting}) quickly reaches its maximum value $1.0$ where all starting 
attempts are accepted. The approach towards the limiting value is of course strongly coupling dependent, and for 
coupling set II the increase is much slower. The number of termination attempts
$N_{TA}$ increases with $A$ as expected, but of course without reaching an asymptotic value. 
While the starting ratio $R_S$ has no finite size effect (as expected), the number of terminating attempts $N_{TA}$ 
shows a mild volume dependence which we attribute to worms closing around the periodic boundaries. It is interesting to note that 
$N_{TA}$ reaches rather high numbers (which again depend on the couplings), which implies that for large $A$ the worm 
might spend a lot of time in the vicinity of its starting point, trying unsuccessfully to close the loop.
 
As expected, the increase of $N_{TA}$ with increasing $A$, implies that also the number of (accepted) steps $N_{steps}$ of the worm 
increases with $A$. This behavior is seen in Fig.~\ref{steps}. Also here we observe a mild finite size effect, 
which again can be attributed to worms that wind around the periodic boundaries. 

Finally, in Fig.~\ref{zero} we show the ratio of sterile worms as a function of $A$. Sterile worms are those worms
where the worm retraces exactly all previous steps and then terminates without changing the flux variables. In other words, 
sterile worms consist only of dangling dead ends. 
One expects that the number of sterile worms goes down as the worms become longer, i.e., as $N_{steps}$ increases, since
for longer paths the probability to exactly 
retrace all steps goes down. Actually the largest fraction of sterile worms are the 2-step worms, where after the starting step the worm is canceled immediately by the second step. It is obvious that these 2-step worms are strongly affected by any change of the terminating probability determined by the amplitude $A$. And indeed we observe that as $A$ is increased (leading to larger
$N_{steps}$) the ratio of sterile worms $R_{SW}$ goes down. However, the decline is not necessarily uniform as is obvious for 
coupling set I shown in the lhs.\ plot of Fig.~\ref{zero}. We furthermore 
observe, that $R_{SW}$ has no finite size effects. This is not a surprise,
since sterile worms that are made only of dangling ends cannot wind and thus do not see the finiteness of the volume. 
The absence of any volume dependence for $R_{SW}$ further supports our interpretation that the finite size effects observed for 
$N_{TA}$ and for $N_{steps}$ come from worms that wind around the periodic boundaries. 

We conclude our analysis of the influence of the 
amplitude parameter $A$ on properties of the worm by studying the actual amount of flux 
that is changed by the worm. For this analysis we define
\begin{equation}  
\Delta_{flux} \; = \; \sum_{x,\nu} | \widetilde{k}_{x,\nu} - k_{x,\nu} | \; ,
\end{equation}
where $k_{x,\nu}$ and $\widetilde{k}_{x,\nu}$ are the fluxes before and after the worm. Note that this definition goes beyond
simply determining the size of the loop ${\cal L}$ that is changed by a worm, but takes into account that a worm could run through 
sub-loops several times thus creating changes of flux larger than $\pm 1$.

We show our results for $\Delta_{flux}$ as a function of $A$ in Figs.~\ref{length} and \ref{change}. In Fig.~\ref{length} 
$\Delta_{flux}$ is normalized by the total number of links $4V$, while in Fig.~\ref{change} we normalize by the number 
of accepted worm steps $N_{steps}$. When normalizing by the number of links (Fig.~\ref{length}) we find that 
$\Delta_{flux}/4V$ increases as $A$ increases and the worms become longer. This is indeed as expected. On the other hand,
when normalizing by the number of worm steps $N_{steps}$ we find that $\Delta_{flux}/N_{steps}$ decreases when $A$ is 
increased and thus $N_{steps}$. This means that long worms are not necessarily more efficient in the sense that they lead to 
larger amounts of flux changed per step. However, a complete efficiency analysis has to take into account also the starting probability
of the worm and depending on the couplings this balance between the two effects might give quite different optimal 
values of $A$ for different couplings. 
The last observation suggests that the role of the amplitude parameter $A$ might go beyond optimizing properties of a 
worm, but that $A$ even could be used to analyze the dynamics of worms by varying their length for a given set of 
couplings and to study how this influences efficiency and other characteristics.

\section{Summary and discussion}
In this article we introduce a general worldline model, where the partitions function has weight factors that live on the links of the
lattice, as well as weight factors 
on the sites of the lattice. The latter are not present in the standard applications the worm algorithm was 
developed for and one has to identify the correct distribution of the site weight factors in the Metropolis decisions of the worm. 
A key complication in the general model is that in a step of the worm the final weights are known only 
at one side of the link. We propose a suitable distribution of the site weight factors in the Metropolis acceptance probabilities
which is independent of any particular form of the site weights. For the starting and terminating steps 
specific Metropolis probabilities are used.  
We analyze the algorithm in detail and give a proof of detailed balance. 

In the starting and the terminating step the site weights are unbalanced, i.e, they appear only in the denominator for the starting step 
and the numerator for the terminating step. We introduce an amplitude $A$ for the starting and terminating steps (there $A$ appears
in the denominator), such that very small or very large 
numbers for the site weights can be compensated for. This allows one to influence the 
starting and the terminating probabilities of the worm and to optimize the average worm length of the algorithm. 

The worldline model we consider is kept rather general, such that a wide class of systems with link and site weights is included. 
Furthermore the key steps of our proof of detailed balance (decomposition of the transition probability into contributions of individual 
worms $w$, identification of the correct inverse worm $w^{-1}$, matching of individual weights for steps of $w$ and $w^{-1}$) 
do not refer to any specific properties of the model, such that our proof can be used as blueprint for even more general models.  

The above mentioned amplitude $A$, which enters the Metropolis ratios for starting and for 
terminating the worm, not only serves to compensate for potentially low starting probabilities, but via adjusting the terminating probability $A$ can also be used to control the length of the worm and to optimize its performance. 
We demonstrate this feature of our worm algorithm in a numerical study of the 4-d relativistic Bose gas. The introduction of such 
a parameter can be generalized to other systems and other variants of worm algorithms in a straightforward way such that the  
performance of worm algorithms can be optimized for different systems and various values of the couplings.

\section*{Acknowledgments} 
\vspace{-1mm}
\noindent
This work was supported by the Austrian Science Fund, 
FWF, DK {\it Hadrons in Vacuum, Nuclei, and Stars} (FWF DK W1203-N16).

\end{document}